\documentstyle[12pt,epsf,epsfig]{article}
 at 10truept
 at 10truept
\font\tinynk=cmr8 at 10truept
\def\be{\begin{equation}}
\def\ee{\end{equation}}
\def\ba{\begin{eqnarray}}
\def\ea{\end{eqnarray}}
\def\bq{\begin{quote}}
\def\eq{\end{quote}}
\def\PL{{ \it Phys. Lett.} }
\def\PRL{{\it Phys. Rev. Lett.} }
\def\NP{{\it Nucl. Phys.} }
\def\PR{{\it Phys. Rev.} }

\def\ltap{\ \raise.3ex\hbox{$<$\kern-.75em\lower1ex\hbox{$\sim$}}\ }
\def\gtap{\ \raise.3ex\hbox{$>$\kern-.75em\lower1ex\hbox{$\sim$}}\ }
\def\gl{\ \raise.5ex\hbox{$>$}\kern-.8em\lower.5ex\hbox{$<$}\ }
\def\roughly#1{\raise.3ex\hbox{$#1$\kern-.75em\lower1ex\hbox{$\sim$}}}

\begin{document}

\thispagestyle{empty}
\begin{flushright}
NYU-TH/99/11/01\\ SU-ITP-99/49\\
hep-ph/9911386\\ November 1999
\end{flushright}
\vspace{1mm}
\begin{center}
\vspace*{1cm}
{\Large \bf MANYFOLD UNIVERSE}\\
\vspace*{0.8cm}
{\large Nima Arkani-Hamed$^{a,b,1}$,
Savas Dimopoulos$^{c,2}$, Gia
Dvali$^{d,3}$
}\\
\vspace{1.5mm}
{\large and Nemanja Kaloper$^{c,4}$}\\
\vspace{0.5cm}
{\em $^a$
Department of Physics, University of California, Berkeley, CA 94530,
USA}\\
{\em $^b$Theory Group, Lawrence Berkeley National Laboratory,
Berkeley, CA 94530, USA}\\
{\em $^c$
Department of Physics, Stanford University, Stanford CA 94305, USA}\\
{\em $^d$ Department of Physics, New York
University, New York, NY 10003, USA}\\
\vspace{2mm}
{\tinynk $^1$arkani@thsrv.lbl.gov,
$^2$savas@stanford.edu, $^3$gd23@scires.acf.nyu.edu,
$^4$kaloper@epic.stanford.edu}\\
\vspace{0.7cm}
ABSTRACT
\end{center}
We propose that our world is a brane folded many times inside the
sub-millimeter extra dimensions. The folding produces many
connected parallel branes or folds with identical microphysics - a
Manyfold. Nearby matter on other folds can be detected
gravitationally as dark matter since the light it emits takes a
long time to reach us traveling around the fold. Hence dark matter
is microphysically identical to ordinary matter; it
can dissipate and clump possibly forming dark replicas of
ordinary stars which are good MACHO candidates. Its dissipation
may lead to far more frequent occurrence of gravitational
collapse and consequently to a significant enhancement in
gravitational wave signals detectable by LIGO and LISA.
Sterile neutrinos find a natural home on the
other folds. Since the folded brane is not a BPS state,
it gives a new geometric means for supersymmetry
breaking in our world. It may also offer novel approach for the
resolution of the cosmological horizon problem, although it still
requires additional dynamics to solve the flatness problem.
Although there are constraints from BBN,
structure formation, the enormity of galactic halos and the absence of stars
and globular clusters with a discernible dark matter component, we
show that the model is consistent with current observational
limits. It presents us with a new dark matter particle and
a new framework for
the evolution of structure in our universe. \vfill
\setcounter{page}{0} \setcounter{footnote}{1}
\newpage
\section{Introduction}

The idea that we live on a brane embedded in a spacetime with $N$
additional large spatial dimensions may provide an alternative understanding of
the hierarchy between the gravitational and electroweak
mass scales \cite{add,add2,aadd}. In this framework,
the fundamental scale of quantum gravity can be as
low as $\sim$TeV, without conflicting with any existing experimental
limits. Then the observed Planck mass, $M_{P} = (G_N)^{-1/2}
\sim 10^{19}$ GeV, where $G_N$ is the Newton constant, is
related to the fundamental Planck scale in $4 + N$ dimensions,
$M_{Pf}$, by Gauss law \cite{add}
\begin{equation}
M_{P} = M_{Pf} \sqrt{ M_{Pf}^N V_N}~,
\label{relation}
\end{equation}
where $V_N \equiv L_1 L_2....  L_N$ is the volume of the extra
space, and $L_i$ is the size of the $i^{\rm th}$ extra dimension.
It is clear that this new picture has dramatic consequences both
for particle physics and cosmology. In addition to shedding new
light on the mass hierarchy problem, in the realm of elementary
particle physics the framework of large extra dimensions has
tremendous consequences for the low energy phenomenology,
accounting for many novel physical phenomena involving fields
living in the extra dimensions, such as for example the new origin
of approximate symmetries from distant breaking \cite{shin,flavorbd,ahsch}, neutrino
masses \cite{addm,dsmirnov}, and flavor generation \cite{flavor}
and violation \cite{flavorbd} near the TeV scale.
It may provide for an alternative
understanding of the unification of gauge couplings \cite{ahdmr}, where
infrared effects in the bulk appear as
ultraviolet effects on the brane \cite{thermal,ahdmr}.
There are also other new effects such as global charge
non-conservation due to production of baby branes \cite{dgabadadze},
and new types of dark matter in the form of super-heavy superstrings \cite{thermal}
or tilted branes \cite{ds,misha}. Connections with string theory have
been examined in \cite{strings}.
Explicit solutions of
Einstein's equations with large compact spaces were
constructed in \cite{ck}.
In fact, the richness
of new phenomena makes a compelling case that theories with
large extra dimensions, and a desert in the bulk, are a fully
consistent alternative to the conventional theories with a great
desert in energy scales, leading to interesting new
phenomenological consequences (for an incomplete list of references,
see \cite{add,add2,ddg,pheno} and references therein).
They are consistent with present observational constraints, as shown in
\cite{add,add2} and confirmed in more detail \cite{observ}.

Cosmological considerations of models with large extra dimensions
confirm that they are a consistent candidate to describe our world.
At the first sight it may appear that cosmology poses very difficult
challenges to these models, requiring a resolution of cosmological
conundrums at energy scales not exceeding $\sim$ TeV.
Furthermore, in addition to the usual cosmological problems such
as the horizon problem and the flatness problem, there may emerge
new problems, involving the possibility of bulk graviton
overabundance \cite{add2,kl}. However, various aspects of
inflation have been considered \cite{kl,dt,ahdkmr,thermal}, leading to the
proposal of several attractive models, such as the
brane inflation of \cite{dt,thermal} and the asymmetric inflation
of \cite{ahdkmr}. There it has been shown that it is possible
for a brane universe to undergo inflation at an early stage,
leading to a universe much as our own at low energies.
Recent studies have considered some of these and
related issues (an incomplete list of references is given in
\cite{ftw,othercos}).

In the present paper we will discuss novel astrophysical and
cosmological implications
of the brane universe with large extra dimensions.
We propose that the brane which we
live on is {\it folded} in the bulk of the extra spatial dimensions,
producing $n_f$ identical braneworlds, or folds. We will refer to
this configuration as the Manyfold. The adjacent folds of a
Manyfold are connected by regions where the brane bends and
its curvature is very large. We will refer to different folds
either as ``folds" or ``branes" completely interchangeably, while the regions
of large curvature where the Manyfold bends will always be singled
out as ``tips" or ``tips of folds" to avoid confusion.
Graphical examples which capture the essential points of these
definitions are given in Figures 1. and 2.

In the Manyfold worlds, observers on different folds can communicate
by exchange of brane-localized degrees of freedom, e.g. electromagnetic waves,
but this requires enormously long times, equal to the sum of their
distances to the nearest tip of the fold.
On the other hand, they can probe objects on adjacent folds
gravitationally, since they are nearby in the bulk, at
sub-millimeter distances.
Hence matter on other folds will appear as dark matter to us.
This dark matter would consist of the {\it very same} Standard Model
particles as our own, localized on an electromagnetically distant part of
the same brane, and with identical microphysical
properties. This property of the folded braneworld
appears counterintuitive at the first glance,
since gravitationally near objects may appear billions of light years away.
However, this is a very natural
consequence of the simple features of the model: {\it the localization}
of the Standard Model particles to a {\it folded} brane which is
embedded in large extra dimensions. A low energy observer on the braneworld
can therefore experience two different minimal distances:
gravitational, defined by minimizing the distances traveled
by bulk particles, and electromagnetic, which corresponds to
minimizing the distances along the brane, and around the tips of folds.

This picture can shed a new light on a number of cosmological
problems, including a new natural dark matter candidate. It
suggests important possible new phenomena which we briefly summarize
here:

\noindent $\bullet$ {\bf Dark matter identical to the normal matter.} This
is a consequence of the construction of the
Manyfold universe, since the dark branes are continuously
connected to ours.
Therefore Manyfolds offer a very elegant new approach to the problem of
dark matter in the Universe.

\noindent $\bullet$ {\bf MACHO stars which do not shine.}
MACHO stars made out of dark matter are identical to ours, but shine along
other folds, and their light has a long way to go before it reaches us.
This can naturally explain why MACHO stars are invisible to us.

\noindent $\bullet$ {\bf Hybrid objects.}
Dark matter on other folds can clump in the
same gravitational wells as the visible matter. This will give
rise to hybrid structures with different luminous and gravitating masses
and deviations from the expected gravitational
effects.

\noindent $\bullet$ {\bf Increased occurrence of collapsing systems.}
The folded universe predicts more collapsing young structures, which can be
subject to observation by gravitational wave experiments.

\noindent $\bullet$ {\bf Neutrino mixing.} If the neutrino masses are generated
due to the mixing with a higher-dimensional bulk fermion \cite{addm},
the folded universe picture selects equal mixing between
the Standard Model and sterile neutrinos.
When there are three folds, the mixing is maximal.

\noindent $\bullet$ {\bf Old light.} The {\it old light}
emitted from adjacent folds could reach us after a sufficiently
long time, carrying information
about the distribution of matter from within, say, our galaxy billions of years ago.

\noindent $\bullet$ {\bf Horizon problem.} The folded universe picture
permits apparently superluminal communication between different segments
of the brane through the bulk.
This could give a
non-inflationary solution of the horizon problem,
if the brane was originally crumpled in a small higher-dimensional box
and later unfolded.

We will also discuss one other important feature of the Manyfold
universe which has very important consequences for both cosmology and
particle physics, although is not directly related to astrophysical issues:

\noindent $\bullet$
{\bf Folded supersymmetry breaking.} A folded brane is not a
BPS object, and may therefore be the source of the observed
supersymmetry breaking in our world. This gives the simplest realization
of the idea of \cite{ds}, where the non-BPS nature of the brane Universe
as source for the observed supersymmetry breaking was first suggested.

The paper is organized as follows. In the next section, we present
the basic idea of the folded universe, describing in simple terms
the setup within which we will carry out our investigation. In
Section 3, we discuss the astrophysical consequences of, and
limits on, Manyfold worlds. Section 4 is devoted to the discussion
of sterile neutrinos which come from the other folds. We will
consider aspects of inflationary cosmology in folded universe
models in Section 5, and explain how inflation can produce
Manyfolds. In Section 6, we will present the horizon problem
from the point of view of superluminal communication through the
bulk. In Section 7, we will turn our attention to the questions
about the stability of the folded brane universe, and folded SUSY
breaking. Finally, we conclude with a summary of
our results.

\section{Bifold}

Here we introduce the basic building blocks of the
folded universe. For simplicity we start with a folded universe
which bends only once in the bulk. From the point of view of
a four-dimensional observer this folded universe, or the Bifold,
contains two branes which are connected by a region where the
master brane bends.
A simple pictorial description of a Bifold is given in Fig. 1.
As we have indicated above, we are using the terminology
where we refer to each side of the Bifold as a brane, or a fold,
completely interchangeably, whereas to avoid confusion we call the
high curvature region the tip of a fold. We will later generalize this
to cases with an arbitrary number of folds, or the Manyfold.

Now consider a 3-brane embedded in the bulk with $N$ extra dimensions.
Although the lowest energy state belongs to a straight infinite brane,
cosmologically such an initial condition is not more likely than any other
chaotic distribution of the brane in the bulk.
For instance, we can imagine that initial
conditions in a spatial region at a high temperature before inflation
where given by a brane which was folded  $n_f$ times per
bulk cross-section. While the orientation of the folds could be
completely random initially, this configuration can then be stretched up
by a subsequent era of early inflation, and converted into $n_f$ parallel
branefold worlds. The difference from other hypothetical brane worlds, which
can also be located in the bulk, is that in the present case the matter
localized on {\it all connected} folds will be {\it identical} by
cosmological evolution and the connectedness of the folds.

\begin{figure}[h,t]
\begin{center}
\epsfig{file=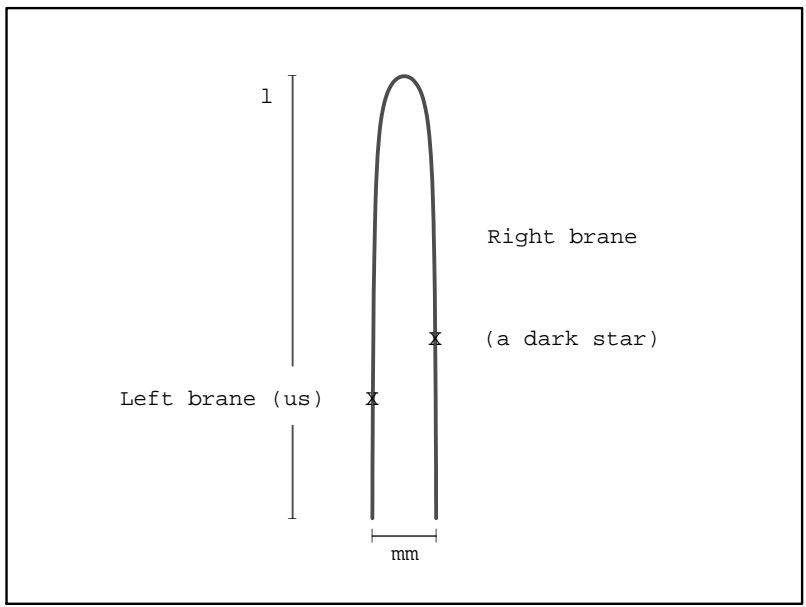,height=8cm}
\caption{\label{fig:Fig1}
{\it The Bifold universe. The distinction of matter between
the Standard Model particles
on the left-brane and the dark particles on the right-brane is not
microphysical but purely geometrical.}} \
\end{center}
\end{figure}

The local universe seen from any of the individual branefold worlds is
rather unusual. The matter localized on the neighboring branes
appears as near-by dark matter, since it is less than a millimeter
away. However, in contrast to the usual dark matter models,
where dark matter particles are completely invisible electromagnetically
at low energies, in this case an observer on a fold could eventually see
the dark matter if she waits long enough. Indeed,
light emitted by the matter on an adjacent fold will
propagate along the brane, get refracted by the tip at
the end of the fold and finally arrive to the observer's side.
She will interpret the long travel time around the tip of a fold by
saying that the source was a very distant object. However, this
object may in fact be just a millimeter away through the bulk!
Thus the dark matter in the halo of our galaxy can be made
out of the very same protons and electrons as we are, which
could be potentially visible, and we may not even know it!

Of course, the key question is how far the nearest tip of our fold
is from us. Obviously, there will be constraints on the
longitudinal size of folds, coming from astrophysical
considerations, and we will discuss them in detail below. For this
purpose, we will imagine that the folded universe consists of
$n_f$ folds, which may have their tips at different distances
from our shadow on each fold. We will first consider the
simplest possibility, which appears to be the most natural in the view
of inflationary cosmology: that the tip of the adjacent
fold is farther from us than
the Hubble radius. Indeed, if the Manyfold has emerged
from a chaotic initial distribution of branes by way
of, for example, asymmetric inflation \cite{ahdkmr}, then
since asymmetric inflation can give may more efoldings along the
brane than is necessary to solve the cosmological problems, the
tips of folds may be at distances many orders of magnitude larger
than the present Hubble horizon. However, we will also generalize this
discussion to include the case when tips
of folds may be at distances smaller than the present Hubble
length. This may come, for example, from branes which were initially extremely
wiggly at short distances, and inflation was not long enough to
blow the wiggles completely outside of our horizon.

Before we proceed, we address a
possible concern one might have about the
Bifold (and Manyfold) world. The first thing one might think
will go wrong with these ideas is that Hubble's law is violated.
For illustration, consider the possibility that we are at a distance
from the nearest tip of our fold which is smaller than the
present Hubble length. Then we can see objects on the adjacent
fold, and in fact can associate two types of positions and velocities
to them (see the points depicted on the folds in Fig 1.)

\noindent (1) in the bulk, denoted $r_{gr}$ and $v_{gr}$ respectively, and

\noindent (2) along the brane, $r_{br}$ and $v_{br}$.

\noindent This means that there are two Hubble's laws in the Bifold
universe, and as long as the relative velocity of the folds is miniscule,
it is easy to see that both Hubble's laws are
satisfied:
\be
v_{gr} = H r_{gr} ~~~~~~~~~~ v_{br} = H r_{br} ~~~~~~
{\rm with} ~~~~ H^2 = \frac{8\pi G_N}{3} \rho_{total},
\ee
where $\rho_{total}$ is the {\it total} energy density.
The relative velocity between the folds can be defined as the difference
of velocities each fold relative to
the background set by primordial gravitational waves, which
give the preferential inertial frame since they are not confined to the folds.
These velocities can be measured by determining the anisotropy
of the gravitational wave background, and can be considered
as a fluctuation in the initial condition of the Bifold,
which can be erased by a stage of early inflation.
Of the two Hubble's laws, the latter is experimentally
checked since we use
the Doppler shift of electromagnetic
photons which travel along the brane.

\section{Astrophysics of Manyfolds}

It is straightforward to generalize the Bifold universe
to the Manyfold universe, with an
arbitrary number of brane folds, as depicted in Fig. 2.
We live on the leftmost brane $L$ and there is a number of ``right-branes"
labeled by the integer $i=1,..., n_f$.
Here we will discuss astrophysical signatures of a Manyfold.
To start with, we will
make two simplifying assumptions:

\noindent (i) The distance $l$ between us and the closest tip
of a fold is much greater than the Hubble length. This was
motivated in the previous section.

\noindent (ii) The bulk fields, such as the dilaton, have
constant $vev$'s. This ensures that the local microphysics
that an observer on any brane sees is identical; i.e. the electron
to proton mass ratio is the same for all observers, on any
of the branes composing the Manyfold.

\begin{figure}[h,t]
\begin{center}
\epsfig{file=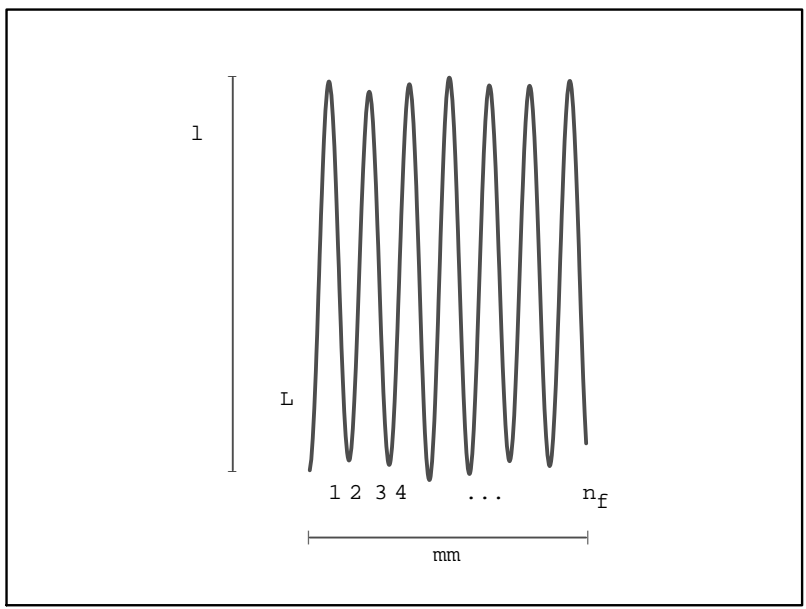,height=8cm}
\caption{\label{fig:Fig2}
{\it The Manyfold universe.
Each side is referred to
as a brane, or a fold, while the region
where they connect is called a tip
of a fold.
We live on the leftmost
brane L, while there are $n_f$ right-branes in the bulk
adjacent to our brane.}} \ \end{center}
\end{figure}

We will later relax each of these assumptions. Also, for
notational simplicity, we assume that the thermodynamic
quantities, such as the temperature $T_R$ and the baryon
number density $n_R$, on all the right-branes are
identical:
\be
T_{i~R} \equiv T_R; ~~~~~~~~~~ n_{i~BR} \equiv n_{BR}
~~~~~~~~~~\forall i = 1,...,n_f.
\label{righttemp}
\ee
Now we turn our attention to astrophysical constraints as well as
possible observable
predictions of the Manyfold universe.

\subsection{Big Bang Nucleosynthesis}

The matter and radiation populating the right-branes (R-branes)
act as dark matter with properties isomorphic to ours. At early
times, its energy density is dominated by the radiation
component. The radiation energy density
is given by $c T^4_R$, where $T_R$ is the
R-brane temperature and $c$ depends on the number of light
species at $T_R$. The success of Big Bang Nucleosynthesis (BBN)
in predicting the abundances of light nuclei demands that
there are no more than one extra species of light
particles at $T_L = T_{BBN} \ltap 1$ MeV. This implies that
the R-branes are colder than our L-brane, $T_R \ll T_L$. In Section
5 we will present an example of how such an asymmetry
between the L- and R-branes can result from a small fluctuation
during inflation.

\subsection{Big Galactic Halos and the Absence of
Hybrid Globular Clusters and Stars}

By the construction of the Manyfold universe, the
matter on R-branes (R-matter) has microphysical properties
identical to the matter on our own L-brane. In particular,
microphysical collisions leading to dissipational
processes have identical cross-sections on the L- and
R-branes. These dissipational processes are
responsible for the formation of condensed
structures in the universe, such as galaxies, globular clusters,
and eventually stars. In the traditional dark matter
models, the dark matter is dissipationless and so it does
not form discernible structures smaller than galactic
halos. In contrast,
R-matter could form smaller structures such as globular
clusters and stars, since it dissipates just like our own
L-matter. Since the L- and R-matter both cluster around
common seeds, which are remnants from an early inflationary era,
this raises a disastrous prospect of forming hybrid stars
or hybrid globular clusters containing both ordinary and
R-matter. Such objects have never been observed. Furthermore,
if the R-matter parallels the evolution of normal baryonic
matter, it could never be an acceptable dark matter candidate.
Since it would dissipate and condense down
to galactic size scales, which are far too small compared
to the size of galactic halos.

A way around these problems is to postulate that the baryon
density $n_{BR}$ on any R-brane is much smaller than our baryon
density $n_{BL}$. This will decrease the rate of dissipation on each of
the R-branes, by lowering the rate at which the R-molecules
electromagnetically collide with each other. Indeed, the ``cooling
time" $t_{cooling}$, that it takes for a plasma to loose a
significant fraction of its energy scales as $t_{cooling} \sim
n^{-1}_B$ \cite{peacock}.
Since our galaxy
formation commenced at a redshift $z \sim 4$, taking
\be
\frac{n_{BR}}{n_{BL}} \ltap \frac14
\label{barratio}
\ee
would ensure that the R-galaxy formation is
beginning just now. Therefore, the R-matter has been behaving
so far as essentially dissipationless cold dark matter,
since it was too dilute to have significantly dissipated its
energy via collisions. Indeed, in the limit $n_{BR}/n_{BL}
\rightarrow 0$, $n_f \rightarrow \infty$, with the
fixed product $n_f n_{BR} \simeq 30 n_{BL}$, the
R-matter would behave {\it precisely} like the ordinary
CDM, critically closing the universe and would form large galactic
halos as a result of gravitational clumping. In this limit,
the R-matter would not significantly clump on shorter length scales,
such as those of globular
structures which are known to be the largest objects
without a significant dark matter content.

Another reason for the difference in cosmological evolution and
structure formation on R-branes is their different light element
abundances.
Because of these differences, BBN on R-branes will proceed quite
differently. Since $T_R \ll T_L$, R-BBN will occur when the energy of
the universe and its expansion is driven by the radiation density
$\sim T^4_L$ residing on our L-brane. The resulting light
nuclear abundances on R-branes will be identical to those of the
standard BBN with a large number of massless species
$\sim c n_f (T_R/T_L)^4$. These scenarios have been studied by
Wagoner \cite{rvw}, and they produce the following outcome.
As the effective number of light species $S = c n_f (T_R/T_L)^4$
increases up to $S_{max} \simeq 10$, the $^{4}He$ abundance grows
until it reaches about $\sim 80 \%$ \cite{rvw}. This occurs because
the universe expands so rapidly that neutrons have no time to decay into
protons, and they combine with protons to form $^{4}He$.
For $S > S_{max} \gtap 10$, the $^{4}He$ abundance drops because the
universe expands so rapidly that $^{4}He$ does not
have time to form.

Subsequent formation of structure in the universe will
depend on the precise chemical abundances of elements.
This is clear since dissipation and consequently the cooling time
will depend on microphysical cross-sections, molecular masses etc.
Although detailed study of structure formation is challenging and
beyond the scope of the
present work, we emphasize that this is in principle a well-defined
program in which, given the ratios
\be
(i) ~~~~~ \frac{T_R}{T_L} ~~~~~~~~~~~~~~~~~ (ii) ~~~~~ \frac{n_{BR}}{n_{BL}}
\ee
one can numerically compute first the primordial nuclear abundances
and subsequently the details of structure formation
inside the R-branes. This should also help us to determine the values
of $T_R/T_L$ and $n_{BR}/n_{BL}$ for which we can comfortably form
R-stars as candidates for MACHOs, to which we now turn.

\subsection{MACHO Star Formation}

An attractive possibility in a Universe of folded branes
is the existence of non-shining R-stars, with masses and properties identical
to those of our stars, with the exception that they are
electromagnetically invisible. Such objects would be natural
candidates for the MACHOs which cause gravitational lensing,
provided of course that there has occurred sufficient dissipation
of R-matter. This requires that the ratio $n_{BR}/n_{BL}$ is not much
smaller than $1/4$, so that the cooling time $t_{cooling}$
is of the order of the Hubble time. Then,
the R-matter dissipation would be significant enough to
allow the formation of R-stars.

There is a tension between
R-star formation, which
places a lower limit on the ratio $n_{BR}/n_{BL}$,
and the absence of hybrid globular clusters or the
existence of big galactic halos,
which favor small values of $n_{BR}/n_{BL}$. It is
beyond our scope to be more precise
on the lower bound on $n_{BR}/n_{BL}$ implied by the
requirement of R-MACHO formation, than to say that
$n_{BR}/n_{BL} \sim 1/4$ would imply that R-dissipation
is commencing now. Perhaps such a value of $n_{BR}/n_{BL}$
is large enough for some small but potentially interesting
and measurable contamination of globular clusters
or even primordial POP III stars by clumped R-matter.
Note that gravitational lensing data favor non-shining
MACHOs of  sub-solar mass, which as R-stars are a natural
consequence of the Manyfolds.

\subsection{Enhanced Gravity Waves at LIGO and LISA}

An important difference between CDM and R-dark matter (R-DM)
is that the latter can rapidly loose energy and violently collapse,
creating gravitational waves. This increases the
number of gravitational wave sources relative to the CDM models. If the
number of violent astrophysical processes happening
today on the R-branes is of the same order as those happening on
our brane, then the total number of gravity wave sources
increases by a factor $\sim 30$ - the ratio of DM to ordinary matter.
In the case where our
distance $l$ from the nearest tips of folds is less
than the Hubble length, the signal of R-sources would be further
enhanced by their close proximity to us by an additional factor
of order of $\sim (l H)^{-1}$.
This leads us to the discussion of constraints and signatures in
Manyfolds where $l \ll H^{-1}$. This corresponds to relaxing the assumption $({\rm i})$
stated at the beginning of this section.

\subsection{Astrophysics Near the Edge of the Fold}

The case when $l < H^{-1}$ offers an exciting possibility that
an object which appears to be very far - electromagnetically -
is gravitationally very close, perhaps even comprising the
dark matter of our very own galaxy! We have touched upon
this possibility in Section 2, and here
we will consider it more closely.

An immediate consequence of the identification of distant
objects with the nearby dark matter is that violent astrophysical
processes which appear to be happening very far away may in
fact have occurred in our galactic neighborhood and have given rise to
much stronger gravitational waves than inferred from the electromagnetic
distance determined by spectral Doppler shifts. As mentioned, this would result
in an enhancement of gravity wave signals at LIGO and LISA of order
$\sim 30 (lH)^{-1}$.

Another potential signature would be an apparent violation
of Newton's force law at large distances, since the
gravitational force would no longer scale as the inverse square
of the {\it electromagnetic distance} of two particles on
different folds. Of course such violations of Newton's force law
are not necessarily undesirable since the very postulate of the
existence of dark matter is forced upon us because of an inconsistency
between Newton's law and what we literally ``see".

However, before embarking on a detailed investigation of these
{\it exotica}, we ought to address possible difficulties with the
case $l < H^{-1}$. This has to do with the requirement that
electromagnetically distant parts of our universe are located in others
folds, where $T_L \gg T_R$, and hence are cooler now. Even more
importantly, they have been much colder during BBN. This
implies that they must have different primordial light element
abundances. In particular, the primordial
deuterium abundance at large distances $d > l$ should be
different than the local abundance. But, the deuterium
abundance has been measured out to cosmological distances
of a fraction of $H^{-1}$ inside Hydrogen clouds
by looking at quasar absorption lines. This at face value suggests
that $l$ is at least a fraction of unity of $H^{-1}$ or larger.
On the other hand, distant $D$-abundance is just one number
(until recently controversial at that) and one could imagine
reproducing it in other ways, such as changing $n_{BR}$
appropriately. This of course could create different observational problems,
namely distant net baryon abundances $n_{BR}$ which are different from the
local one $n_{BL}$. Again, such variations in $n_{BR} \ne n_{BL}$
may be acceptable at cosmological distances.
It seems that the safest choice is to take $l$ to be larger
than a factor of $H^{-1}$, although we have not actually found
a concrete limit but possible problems with smaller $l$'s.

\subsection{Profiles in the Bulk}

If bulk fields do not vary through the bulk,
microphysics on each fold is identical. However,
bulk fields generically
can change in the bulk (see e.g. \cite{chs} for a review),
developing a non-trivial profile.
For example, branes provide sources for the moduli,
which therefore develop nontrivial profiles in the bulk. Due to Gauss Law,
typically these profiles behave as $a + b\exp(-rm)/r^{N-2}$
at large distances,
where $m$ is the mass of the bulk field, satisfying a very
weak constraint $M_{Pf} > m > 10^{-2} eV$, where the lower
bound comes from the fifth force searches.
If either $m$ or $N$ is large, the profile will rapidly approach a
constant value far from the source brane. Otherwise for a small mass
and e.g. $N=2$, the field may even approach a logarithmic shape.
An immediate consequence
of having a non-trivial bulk field profile
is that since the dilaton $vev$ sets the values of gauge couplings on the brane,
different copies of the Standard Model on each folds will have different
strengths of couplings depending on where they are in the bulk.
This produces a natural asymmetry between branes, and can give rise to different
cosmological evolution of different folds even if the initial temperatures after
inflation on different folds were equal.

Indeed, if for example the electron weighs as much as the proton
on the R-branes, then all R-atoms will have very small dipole and multipole
moments $\sim (e/m_p)^\alpha$, and so R-matter will be essentially
dissipationless, just like CDM. Another possibility is that the
$u$ quark may be heavier than the $d$ quark,
and the lightest stable baryon is the neutron, impeding the formation of R-atoms.
Still another possibility is that just the
value of the canonical dilaton field is varying in the bulk, leading to
an overall scaling of couplings.

Since the formation of structure and stellar evolution are so sensitive
to the masses and couplings of elementary particles, even a small variation
can lead to drastic changes and therefore makes them difficult to track down
in detail. Because there is no well-motivated and concrete model
with a solution having specific bulk-varying fields, we will continue to
study the simplest possibility of constant bulk fields. In fact, the hypothesis
of constant bulk fields by itself and on very general grounds necessarily
implies the existence of multiple identical copies of our brane,
without the need for a physical connection between the branes. This
is because the constancy of the bulk fields requires the simultaneous
presence of $D$-branes (positive sources) and coincident Orientifolds
(negative sources). The latter can only occur in multiple copies
in compact spaces. In simple cases, such as toroidal compactifications,
they come in $2^N$ identical copies, where $N$ is the number of new compact
dimensions. In this case, each copy would support locally identical
particle physics although there is no direct physical contact between
different branes, but the phenomenology would still be the same as
in the example of a folded brane, subject to identical
astrophysical and cosmological constraints.

To conclude this section, we point out that our discussion shows
that mirror models with mirror
particles isomorphic to ours, as for example those of \cite{mirror},
are excluded on simple grounds if they
compose more than $1/30$ of the critical density. They dissipate
more than ordinary matter and would form halos smaller than
our galaxy. They would also strongly mix with matter on shorter
length scales leading to hybrid globular clusters of hybrid
primordial stars. To avoid such disasters, one would have to
postulate $\sim 100$ identical mirror models.
Alternatively, postulating an asymmetry between our world and
the mirror world is possible, and can be generated naturally
in Manyfolds with profiles in the bulk. In that case, the mirror particles \
can have different microphysical properties, and evade the astrophysical
constraints with employing fewer copies.

\section{Sterile Neutrinos From Distant Folds}

The Manyfold universe allows for a very peculiar neutrino mixing
for the models of neutrino mass
generation {\it a \'la} \cite{MSW} employing the higher dimensional
mechanism suggested in \cite{addm}. Rather then reviewing the details,
here we will consider how this picture changes in the case of the
Manyfold universe.
Consider first the other side of the fold across our brane.
The ``image" of our left-handed neutrino
becomes a right-handed particle. This is due to the fact
that the brane changes its orientation. As a result, all the localized zero modes
flip their chirality, where the left-movers become right-movers and visa versa.
This reflects the fact that a brane becomes an anti-brane
upon folding (see the discussion in Sec. 7).

To see this in more detail, we consider an explicit example of a kink
(anti-kink) brane (which we will discuss in more detail in
Sec. 7). A kink (anti-kink) brane is supported
by a scalar field which interpolates between two degenerate and disconnected
vacua, $\phi = \pm v \tanh(\lambda v x_5)$, where $\lambda$ and $v$ are its
coupling constant and $vev$, respectively.
Let now ${\cal N}$ be
a five-dimensional
Dirac fermion coupled to this scalar and let us consider Dirac equation
for this fermion in the kink (or anti-kink) background:
\begin{eqnarray}
\left( i\Gamma^{\mu} \partial_{\mu} + i\Gamma_5\partial_5
- \phi \right ){{\cal N}} =0~.
\label{fermion}
\end{eqnarray}
It is well-known that this equation admits only one localized
chiral mode
\begin{equation}
{{\cal N}}_{L,R} (x_5,x_{\mu}) = {{\cal N}}_{L,R}{\rm e}^{\pm\int_0^y\phi_(y)dy},
\label{normnu}
\end{equation}
where the chirality is defined by the boundary conditions. The reason is that
in infinite extra dimensions there is
only one normalizable mode for a specific choice of the background,
as is clear from the exponential factor in (\ref{normnu}).
Alternatively, when the bulk is compact, only one of the modes
is dynamical while the other effectively decouples.
Therefore, if the brane localizes
the left-handed fermion, the other side of the fold can only localize the
right-handed one.

In the case when a brane and an anti-brane are connected,
the right-handed neutrino which is localized on the other side can be thought of
as our left-handed neutrino which changes chirality
after refraction around the tip of our fold.
The impact of this observation on the structure of the neutrino mass matrix
is that now the zero mode of the left-handed bulk state ${\cal N}_{0L}$
is not decoupled anymore, as in \cite{addm},
but mixes with the right-handed image
$\nu_R$ on the anti-brane, and becomes massive.
The mass matrix can be written as:
\be
\bar{\nu}_L M {\cal N}_R~,
\ee
where ${\nu}_L^T \equiv (\nu_L, \tilde{\nu}_{1 L},
\tilde{\nu}_{2 L} ... )$ and
${\cal N}_R^T =  ({\cal N}_{0 R},
\tilde{{\cal N}}_{1 R}, \tilde{{\cal N}}_{2 R} ... )$;
the  modes
${\cal N}_{0 L}$, $\hat {\cal N}_{n L}$ $\hat {\cal N}_{n R}$
decouple  from the system. The mass matrix
$M$ for any $k+1$ low-lying states is
\be
M = \left(
\begin{array}{cccccc}
0 & m_D & \sqrt{2}m_D &  \sqrt{2}m_D  & ...   & \sqrt{2}m_D \\
m_D & 0 & 0 & 0 & ...   & 0 \\
\sqrt{2}m_D  & 0 & \frac{1}{R}  &  0  &  ...  & 0   \\
\sqrt{2}m_D & 0 & 0   &  \frac{2}{R} & ...   & 0   \\
... & ... & ...  & ...  & ...   & ...  \\
\sqrt{2}m_D & 0  & 0 & 0   & ...   & \frac{k}{R}
\end{array}
\right)~.
\ee
Now, it is straightforward to generalize
this structure for an arbitrary number of folds.
We refer to the L-brane, that we reside on,
as the $0^{\rm th}$ brane,
and count the R-branes in the usual fashion
(see Fig. 2). With this enumeration, the electroweak
doublet neutrinos
which are localized on even-numbered branes
$n_{brane} = 0,2,4,...,2k$
will be left-handed for observers on the L-brane,
$(\nu_{L}^{(0)}, \nu_{L}^{(2)}, ... \nu_{L}^{(2k)})$,
just like in the Standard Model,
while the neutrino doublets
localized on odd-numbered branes $n_{brane} = 1,3,...,2k+1$
will be right-handed,
$(\nu_{R}^{(1)}, \nu_{R}^{(3)}, ... \nu_{R}^{(2k+1)})$.
Therefore we see that the zero mode of the right-handed bulk
neutrino ${\cal N}_{0R}$ will pair up with all the
left-handed brane modes with {\it exactly equal} mass terms:
\be
m_D\bar{{\cal N}}_{0R}(\nu_{L}^{(0)} + \nu_{L}^{(2)} + ... + \nu_{L}^{(2k)}),
\ee
and similarly, the left-handed zero bulk mode ${\cal N}_{0L}$ will pair up
with the right-handed brane states.
This implies that an observer on our L-brane will perceive
all the states $n_{brane}\neq 0$ as sterile neutrinos. Thus this scenario
predicts the equal mixing between our $\nu_L$ and all sterile neutrinos,
with the mixing angle determined only by the number of folds.
How many modes become massive depends on the
number of higher dimensional species in the bulk.
In the simplest case of a single bulk neutrino ${\cal N}$, the only
massive combination is mostly
\be
(\nu_{L}^{(0)} + \nu_{L}^{(2)} + ... + \nu_{L}^{(2k)}),
\ee
while the others remain massless. Note that in the case of
Threefolds, the mixing mechanism is maximal. Otherwise, maximal
mixing may be generated by profiles in the bulk, where the mixing
angle would depend on the bulk field in addition to the number of
folds.

Since this picture predicts a number
of new, light, sterile neutrinos with equal
mixing, the nucleosynthesis bounds discussed in \cite{dsmirnov}
should obviously be reconsidered.
The reason is that now $\nu_L^{(2k)}$ can be produced
by neutrino oscillations. Hence the revised nucleosynthesis bound reads
\cite{ns}
\be
\Delta m^2 < \frac{3 \cdot 10^{-5} {\rm eV}^2 } {\sin^4 2\theta}~.
\ee
Since in our case $\sin^4(2\theta) \sim 1$, this is essentially
a bound on $m_D$, giving $m_D \ltap 10^{-2}$eV.

\section{Initial Conditions for the Manyfold}

We have seen in the preceding discussion that one of the main
constraints on the Manyfold universe is that the baryon number
density on our L-brane is substantially larger than on all the
R-branes. At first sight, this may appear as a serious fine
tuning problem. However, we will argue that, in fact, inflationary
cosmology may naturally produce such an asymmetry provided the
inflaton is a wall-localized field.

In order to simultaneously produce the
four-dimensional Planck mass $M_{P}$ according to
(\ref{relation}) and
to generate the density
perturbations in the range required by COBE,
inflation in the scenarios with a very
low Planck scale
should occur before the extra dimensions are stabilized.
A very natural candidate for the bulk inflaton
is the radius modulus (radion) \cite{ahdkmr}.
A difficulty with any generic scenario of bulk-field-driven inflation
is that it usually produces a Universe
with a cosmological moduli problem, because after the exit from the
early stage of inflation, the universe is dominated by coherent
oscillations of a weakly-coupled modulus.
Another possible objection is that bulk inflaton could reheat
the whole bulk, which could lead to the overproduction of bulk gravitons \cite{add}.
In the asymmetric inflation scenario however the latter problem
is easily avoided, as was discussed in \cite{ahdkmr}. However,
these problems can be solved by
a secondary stage of inflation, which appears to be
necessary regardless of the early cosmological
history \cite{dt,thermal}.

A viable scenario can then unravel as follows.
First, the asymmetric
inflation can start in a region of the bulk where the brane
is very curved. Then, rapid expansion in the longitudinal
directions will stretch and parallelise the folds of the brane,
producing, generically, a Manyfold whose tips are at distances
much larger than the cosmological horizon at late times.
If the period of asymmetric inflation is short, then the tips may
be at distances comparable to the scale of the Hubble horizon.
This stage of early
asymmetric inflation however would produce the same matter energy density
on all folds, and a Manyfold dominated by its radion modulus.
A subsequent stage of wall inflaton-driven inflation can then
produce the situation where
a single fold ends up with the highest reheating temperature, and
the largest baryon number density, whereas the remaining folds turn out
to be much colder. Indeed, let
$\Phi$ be a scalar field localized on the brane,
with the potential of this field $V(\Phi)$ which has a region
flat enough to satisfy the slow-roll conditions. If $\Phi$ is
initially displaced from the minimum, it can drive inflation after the
stabilization of the extra dimensions. Since it couples to the
Standard Model particles on the brane with non-gravitational,
renormalizable couplings of the form
\begin{equation}
g\Phi\Psi\Psi.
\label{inflatoncoupling}
\end{equation}
after inflation ends, $\Phi$ will
begin to oscillate around the minimum and decay
into the fields $\Psi$. If the
coupling constant $g$ is large enough,
the decay will be rapid enough producing a
universe with normal relativistic particles, no
moduli problem, and the reheating temperature
high enough for nucleosynthesis to proceed.

When there is $n_f+1$ similar folds, then each fold supports
an inflaton which couples to a replica of the
Standard Model on its fold according to Eq.
(\ref{inflatoncoupling}) and only {\it gravitationally} to the
copies of the Standard Model residing on the other folds.
Since gravity can propagate through the bulk, each individual
inflaton can simultaneously drive inflation on all of the branes,
but can reheat efficiently only {\it its own} brane.
Since the wall inflatons start out with different initial values,
which can in fact be set up naturally by a stage of
asymmetric inflation of \cite{ahdkmr}, which stretches the spatial
fluctuations of wall fields to very large distances.
Hence the fluctuations whose wavelength is comparable to the size
of the fold will be seen as wall fields lying at different
distances from the minimum of the potential.
Therefore they will reheat their respective folds at different times.
In general, all the reheating energy on the folds where the inflaton stepped
out of the slow roll early will be redshifted away by inflation
driven by inflatons on the other branes, until the last inflaton relaxes to its
minimum. The observable energy density in the Universe will come predominantly
from this last reheating, and we denote the inflaton responsible for this as
$\Phi_{our}$.
As we discussed above, it couples directly to fermions on its own brane
\begin{equation}
g\Phi_{our}\psi_{our}\psi_{our}
\end{equation}
and gravitationally to the fermions on the other branes
\begin{equation}
g'\Phi_{our}\psi_{other}\psi_{other}.
\end{equation}
Here $g'$ is a dimensionless ratio of the scalar $vev$ of $\Phi_{our}$
to the Planck mass, mimicking gravitationally suppressed couplings,
$g' \ll g$.
Although the energy density released on ``our''
brane $\rho_{our}$ is controlled by $g$, while the
fraction released on any other brane will
be suppressed by inverse powers of $M_{Pf}$,
the total energy density released on other branes may
still exceed the energy density $\rho_{our}$ on our brane,
since the multiplicity of channels can over-compensate
gravitational suppression, depending on the ratio $g/g'$
and the number of folds. Assuming that this in
fact happens, although the baryons on any other brane are fewer
in number, their integrated energy density dominates over the energy density
on our fold. This explains our choice of terminology, where we refer to the
fold with the longest-lasting inflation as ``ours",
since it is the only brane that could support life so far!

A very simple model of wall inflation is the brane inflation
\cite{dt}, where the inflaton is the inter-brane separation whose
variation is generated by the motion of branes in the bulk.
Recalling that the branes are
the $D$-branes of string theory, the brane on which we
reside is really a stack of $m$ parallel $D$-branes,
generating an unbroken $SU(m)$ gauge symmetry on the brane.
This should not be confused with the $n_f+1$ folds, which are
all parts of this same stack,
supporting the geometric ``mirrors"  with precisely the same $SU(m)_n$
symmetry. In the early Universe, the slices can be randomly displaced
and start falling towards each other.
If $\Phi$ is identified with the field
which parameterizes the inter-brane distance $r$,
$\Phi = M^2_{Pf}r$, then the
motion of branes is equivalent to the rolling of the scalar field $\Phi_m$
in the adjoint representation of $SU(m)$, whose vacuum energy
drives inflation \cite{dt}. The inter-brane potential
is sufficiently flat to drive inflation, since it is
generated by the bulk closed string exchange, which
gives rise to terms with
inverse power-law behavior at distances $r \gg M_{Pf}^{-1}$.
When SUSY is broken, a good approximation for the potential is \cite{dt}
\begin{equation}
V(r) = M^4{a(M_{Pf}r)^{N -2} + b_ie^{-m_ir}  - 1 \over (M_{Pf}r)^{N
-2}}~,
\label{thepotential}
\end{equation}
where $M$ is the brane tension, $a, b_i$ are some constants,
and $m_i$ are masses of the heavy bulk modes
(e.g. RR or dilaton fields, if these are massive).

Hence a natural scenario for brane inflation in the Manyfold
universe is to take a randomly distributed gas of branes as the
initial condition, produced by asymmetric inflation \cite{ahdkmr}.
These branes can fall towards their equilibrium points,
and during their motion longitudinal inflate exponentially.
The branes reheat when they collide, but the
reheating products redshift away as long as any pair of
branes remain in motion, straggling behind the rest.
This very last moving branes will drive the last stage of
inflation and be responsible for the final stage of reheating,
ending up as the branes with the highest baryon content. These are
the branes where we can exist. After reheating, subsequent
cosmological and astrophysical history will unravel according to
the discussion in the previous sections.

We should stress here that the microphysical coincidence of the dark matter
particles with the Standard Model ones is a direct consequence of
the Manyfold world which emerges from inflation.
This is one of the key differences of our framework from any
conventional models with mirror dark matter \cite{mirror}.
Here we are {\it not} postulating
the existence of any otherwise arbitrary ``mirror" worlds which happen
to accommodate the same microphysics as our world. As we have discussed above,
the mirror worlds are {\it created} by the
cosmological dynamics, evolving from the initial conditions
which are set up after inflation
and are parts of our own brane.

\section{The Horizon Problem}

In a Manyfold universe, different folds can communicate at rates which appear
superluminal to a brane-localized clock, composed of the Standard
Model particles. Such signals must propagate through the bulk,
making short-cuts through the extra space.
Hence different folds could correlate faster through the bulk than by
the emission of light pulses traveling along the brane.
However these effects do not violate causality of the theory.
An observer on the fold, ignorant of bulk gravity,
might mistake the information carried by
bulk gravitons as a causality violation, but this is merely an illusion,
since once all degrees of freedom are accounted for, the theory
is local and causally well behaved.

This offers a non-inflationary possibility for solving the
cosmological horizon problem.
Indeed, consider
a 3-brane large enough to contain the comoving volume of
our observed universe, but initially
very densely crumpled to fit inside a small region of the bulk.
The reason the brane can fit inside such a small volume is that
it is a smooth manifold of lower dimension, and hence it occupies
only a small portion of the bulk (the lower bound on the bulk
volume into which a brane can fit
is set by the thickness of the brane $d$ and its linear
dimension $L$, and is $V_{bulk} \gtap L^3 d^N$).
The brane can dynamically unfold later.
By ``unfolding" we mean that while most of the crumpled parts
will straighten out, there will still remain a number of large parallel
folds with different temperatures and the size equal to or larger than the
comoving horizon scale. If the duration of unfolding
$\tau$ is sufficiently long so that different parts of
a fold come into causal contact, the horizon problem would be
solved, since the background radiation can thermalize.
Thermalization can proceed via two different scenarios, the
``gravity-mediated" scenario, where the information is conveyed by bulk gravity,
and the ``gauge-mediated" scenario, where the information carrier is
a brane-localized gauge boson, e.g. electromagnetic waves.
The main difference between the latter case and the conventional
cosmology is that due to brane dynamics,
the evolution of the causal structure of the crumpled brane
is halted by brane dynamics for the duration $\tau$ of this stage,
permitting the regions of a fold to thermalize.

The process for unfolding the brane is clearly the main
ingredient of the solution. A simple dynamical model which
provides insight into brane dynamics may
be constructed as follows.
At large distances the dynamics of a 3-brane is governed by the
Nambu-Goto action
\begin{equation}
S_{brane} = \int d^4x \sqrt{-g} M^4
\label{minarea}
\end{equation}
where for simplicity we have taken a 3-brane moving in some
fixed bulk background.
The coordinates $x$ parameterize the world-volume of the brane,
$g_{\mu\nu}$ is the induced metric on the brane, and
$M^4$ is the brane tension, which we
treat as a free parameter.
We will ignore gravitational back reaction since
the main effect of gravity is to cause the brane to collapse,
and this can be consistently modeled by the action (\ref{minarea})
alone. Namely, dynamics of the brane is defined by extremizing this action.
If we consider an almost flat, infinite brane, then the solutions
to the extremal problem correspond to the brane moving with
an arbitrary constant velocity, in perfect analogy
with the motion of a free relativistic particle. However,
if we consider a spherical brane,
the action (\ref{minarea}) contains an instability
which leads to brane shrinking. This is because of the
curvature of the brane: for example, consider a
spherical 3-brane of radius $R$ moving in a flat $5D$ spacetime.
In this case, the action (\ref{minarea}) reduces to
$S = - 2\pi^2 M^4 \int dt R^3 \sqrt{1-\dot R^2}$,
and the ensuing Euler-Lagrange equations reduce to a
single equation
\be
\dot R = \pm \sqrt{1- C R^6}
\ee
where $C$ is a constant of integration.
This shows that starting with any initial velocity,
the brane can expand up to the maximum radius $R_{max} =
C^{-1/6}$, after which it collapses back to zero size.
The instability will persist for any
sufficiently crumpled brane. On the other hand, in order
to solve the horizon problem the brane must be crumpled into
a very small bulk volume, where the instability would
try to prevent the
brane from ever unfolding.

The brane can resist the collapsing instability if
there is conserved nonvanishing particle number (i.e. chemical potential)
on the brane.
Indeed, let there be a thermal
bath of particles localized on the
brane, e.g. the Standard Model
fermions, at a high temperature. At first we can consistently neglect
the cooling by bulk graviton emission, and assume that the
entropy on the brane is conserved. Consider now a segment of a
crumpled brane, and again for simplicity let the segment be spherically
symmetric with the radius $R$. Ignoring the motion
of the brane through the bulk to the lowest order,
roughly the quantity which governs dynamics is
\begin{equation}
F \sim (M^4 + T^4)R^3,
\label{fe}
\end{equation}
where $T$ is the temperature. In the absence of particles on the brane,
the segment would collapse, in agreement with the results above.
However, if $n \sim T^3 \ne 0$, the gas has
pressure on the brane, which opposes
its collapse. Indeed, minimizing (\ref{fe}),
subject to the condition of constant entropy
\begin{equation}
TR = {\rm constant},
\end{equation}
we get the minimum size for the curvature
\begin{equation}
R_{min} \sim \frac{1}{M}
\label{radius}
\end{equation}
and the corresponding temperature
\begin{equation}
T_{average} \sim M.
\end{equation}
Therefore we see that in the absence of gravity a crumpled brane
filled with a heath bath at some high temperature $T_{initial}$
will reach a state of minimal radius and {\it uniform} temperature.
This means, the brane will thermalize spontaneously by evolving
to this state! At this point the brane becomes effectively
{\it tensionless}, and it does not collapse any further.
We note that this simple
toy model is very similar to a star being stabilized
by radiation pressure against gravitational collapse.
Indeed, there too the temperature of a star becomes essentially homogeneous
with the exception of perturbations. To push the similarity a little
further, we could compare the process of unfolding the crumpled brane
with a star going supernova.

In a more realistic situation, the brane
will cool by dynamical effects.
There are two main model-independent
channels for cooling: 1) evaporation the bulk gravitons \cite{add}
or baby branes \cite{dgabadadze} into the bulk,
if it is colder than the brane; and
2) cosmological expansion, when gravity is turned on.
We will ignore the former effect and focus only on the latter,
which corresponds to adiabatic cooling\footnote{Evaporation of baby branes
requires some clarification in the present context.
Normally this process is exponentially
suppressed, $\sim {\rm e}^{-{E/T}}$, where $E$ is of the order of the energy of
a baby brane \cite{dgabadadze}. In the present case, since the brane
is effectively ``tensionless" once the equilibrium is reached,
$M \sim T$ and the production of tiny baby branes may be unsuppressed.
However, note that even if it is unsuppressed, this process will not
significantly alter the entropy of the
brane, since a baby brane of the size $R_{baby}^3$ can only take entropy
$\sim T^4R^3_{baby}$.}. The idea here is that the
expansion of the universe will only begin
to influence the evolution after a time $\tau$,
needed for the Universe to expand
by a factor $L/D$. If this coincides with the time needed for a
tensionless crumpled brane of size $V$ to unfold, during which
the temperature of the brane changes very little,
and the brane remains thermalized so that
immediately after this epoch the Universe can expand according
to the usual four-dimensional FRW model.
Hence $T\sim M$ is the {\it normalcy} temperature of
the universe, defined in \cite{add} as the temperature
below which the cooling by bulk graviton
emission is negligible. Numerically, to make sure that BBN can proceed
normally, we must require that $T\sim T_* > {\rm MeV}$.

To gain a more quantitative description of these phenomena,
we consider a 3-brane of volume $V_3 = L^3$ which is
crumpled into a small fraction of the bulk, of volume $V_{3 + N} = D^{3 + N}$.
If the box size is much smaller than the longitudinal size of the brane, $D \ll
L$, we can solve the horizon problem, provided that
$L$ is equal to or larger than
the comoving horizon size today,
and $D$ is of the order of the Hubble length
at the time scale set by the brane temperature. Then
different parts of the 3-brane can thermalize
if either of the
following two conditions holds:
\ba
&&L\sim L_*,~~~~~~~~~~{\rm and}~~~~~~~~~~ D \sim H^{-1} ~~~~~~ {\rm or}
\nonumber \\
&&\tau \sim L_*,
\label{tw}
\ea
where $L_*$ is the comoving scale of the present Hubble size evaluated
at the temperature $T_*$ in a {\it conventional} FRW Universe,
and $H$ is the {\it effective} Hubble parameter at that time.
These two conditions correspond to gravity-mediated and gauge-mediated
scenarios respectively.

Unfortunately the conditions (\ref{tw}) are very hard to satisfy.
Indeed, the minimal amount of energy of a crumpled brane is
$\sim L_*^3T_*^4$. This must be squeezed into a
higher-dimensional box of volume $V_{3+N} \sim D^{3 +N}$.
Therefore the higher-dimensional Hubble parameter
which corresponds to the energy density $\rho \sim L^3_*
T^4_*/D^{3+N}$ is $H^2 = L_*^3T_*^4 /(D^{3 +N}M_{Pf}^{2
+N})$. But in fact let us assume for simplicity that
the size of extra dimensions is stabilized (it can be easily
shown that relaxing this condition
does not alter the conclusions).
This is self-consistent only if $H^{-1} > R$,
otherwise the bulk would fragment into many disconnected
regions of Hubble size. Since the transverse size of a region
is then bounded by $\sim R^N$, the total volume is
transmuted into $D^{3+N} \rightarrow D^3R^N$, and we get
\begin{equation}
H^2 = {L_*^3T_*^4 \over D^3M_P^2} = {L_*^3\over D^3}H_*^2
\end{equation}
where $H_*^2 = {T_*^4 \over M_P^2}$ is the corresponding
normal four-dimensional Hubble parameter in an
FRW Universe at the temperature $T_*$,
and we have used Eq. (\ref{relation}) to eliminate $M_{Pf}$.
Since $D$ must be at least as large as $H^{-1}$, the Hubble
parameters must satisfy
\begin{equation}
H^{-1} = L_*(L_*H_*)^2.
\end{equation}
In a $4D$ FRW universe $L_*H_* > 1$, and hence the above expression can
hold only if $T_* \sim 3^o$K, meaning that
we cannot fit an arbitrarily large brane
into a given higher-dimensional Hubble volume: pushing the brane in
makes the volume smaller! This is because the brane must carry entropy
necessary to protect it against the instability which causes collapse.
A loophole in the above argument could be
provided by a brane which is effectively tensionless,
and with only a small amount of entropy deposited on it.
If this is the case, then it may be possible to thermalize the
crumpled brane at a higher temperature, however such solutions
are not known at present. The second condition (\ref{tw}) requires
a slow unfolding of the brane. In
the brane unfolds due to the expansion of the universe,
the time it takes is
\begin{equation}
\tau \sim  1/H_*.
\end{equation}
Again, if this is to satisfy the condition (\ref{tw}), the temperature must
be $T_* \sim 3^o$K, which again is too low.

Both of these mechanisms point toward the
same value of $T_*$, but unfortunately at temperatures far below MeV.
Hence while it is conceivable that the horizon problem may be solved,
at present this mechanism does not give a universe
which is homogeneous at temperatures hot enough for BBN to
proceed unhampered. Furthermore, this process also does not
address the flatness problem, which would become
acute when gravity is turned on and general initial conditions
in the bulk and on the brane are allowed.
Namely, the temporary resistance of the brane to collapse
generated by entropy deposited on it might suggest that
even flatness can be addressed. However, when gravity is turned
on, it will lead to stronger attraction, since the energy
carried by the gas on the brane will add to the gravitational binding,
and possibly disrupt the balance which kept the brane from imploding.
Nevertheless, it is
interesting to note that at least the horizon and the flatness
problems might be somehow decoupled from each other.
Further, the difficulty in solving the horizon problem
with the crumpled brane scenario
above arises in the few simple cases which we have considered.
If there are certain binding interaction, which would
slow down unfolding, the problem could be
avoided. It would be interesting to see if such a mechanism could be
provided by long strings which can stretch between different
parts of the brane.

\section{Stability of the Folded Brane and Folded SUSY Breaking}

When a brane folds, its two sides behave as a brane and an anti-brane
(ignoring the tip of the fold).
For instance if the brane in question is a $D$-brane, the two sides will
carry opposite RR
charge\footnote{We thank Nathan Seiberg and Massimo Porrati
for useful comments on this issue.}.
Hence in general the two sides can attract, and could in fact annihilate.
This questions the stability of the Manyfold universe,
and requires some clarification.
The brane-anti-brane correspondence between any two adjacent
folds can be simply understood by noting that
the world-volume element to which an RR form couples,
\begin{equation}
S_4 = \mu\int d\sigma C_{(4)},
\end{equation}
changes sign after folding. This happens because the folding
corresponds to changing the orientation of the brane,
which is induced by the reflection of one of the coordinates on
its world-volume. Another way to understand this is to
note that the charge of a closed brane must be zero.
Indeed, the charge of a $D_p$ brane embedded in a spacetime with
$N$ transverse dimensions can be estimated by
evaluating the integral from the dual of a RR field strength
around a closed $N-1$ dimensional surface
which encircles both folds in the transverse space
\begin{equation}
\int_{S_{N-1}}*F_{(p+2)}.
\label{rrcharge}
\end{equation}
For a closed brane, this surface can be deformed and shrunk to a point,
which shows that the integral Eq. (\ref{rrcharge})
must vanish and so the RR charge must vanish too.

Therefore this brane-anti-brane pair is not a BPS state any more,
because now the force mediated by the exchange of the RR gauge bosons adds
to the dilaton and graviton attraction
instead of compensating them. The system becomes unstable,
with a tendency to collapse.
On the other hand, the non-BPS nature of the folded Universe may even be
welcome, since it would break supersymmetry in the real world.
This is in the spirit of the approach of Ref. \cite{ds}, where it has been
suggested that non-BPS walls may be the source of the observed supersymmetry breaking.
However, to realize this possibility in the present context, one must develop
ways to stabilize the Manyfold structure, cancelling the overall
attraction between the folds.
In this section, we consider examples where a brane-anti-brane
pair can be stabilized at some finite separation
due to additional bulk states in the theory.

Consider a simple field theoretical example of a brane embedded in
a space with a single transverse dimension. Such a brane can be modeled
by a domain wall generated by a
real scalar field with a discrete symmetry breaking potential:
\begin{equation}
V = \frac{\lambda^2}{2}(\phi^2 - v^2)^2.
\label{pot}
\end{equation}
This system has topologically stable solutions,
exemplified by the kink/anti-kink configurations:
\begin{equation}
 \phi =  \pm v {\rm th} (\lambda v x_5)
\label{kink}
\end{equation}
where $\pm$ refers to the kink and the anti-kink respectively.
Viewed separately, each of these solutions is stable due to non-vanishing
topological charge. However, a kink-anti-kink
pair has vanishing total charge and therefore is unstable.
The configurations with vanishing total charge can be stabilized if there are
other bulk fields. For instance let us introduce another real
field $\theta$ defined modulo $2\pi$: $\theta$ can be thought of as the
phase of a complex field with a nonvanishing $vev$.
This means that the scalar field potential can now be written as
\begin{equation}
 V =  {\rm cos}\theta(-a - b\phi^2) + \frac{\lambda^2}{2}(\phi^2 - v^2)^2,
\label{crosscouplings}
\end{equation}
where we could have added many other terms compatible with the symmetries
$\phi \rightarrow -\phi,~~\theta \rightarrow \theta + 2\pi$, without altering
any of our conclusions below. However, for the sake of simplicity
we will restrict our attention only to the potential of the form
(\ref{crosscouplings}), since it is sufficient for the purposes of the argument.
Now in the presence of the field $\theta$ there arise two types
of topological defects, as can be seen by simple topological considerations:
kink (anti-kink) given in (\ref{kink}) (where $v$ should be renormalized to
account for the additional terms in the potential) and
sine-Gordon solitons, the simplest of which are those
along which the field $\theta$ changes by $2\pi$ when
$x_5$ goes from $-\infty$ to  $+\infty$.
Let us compactify the fifth dimension on a circle and consider the case
with two solitons and a kink-anti-kink pair. Note that since
the periodicity of $\theta$ is consistent with periodic boundary conditions,
there may be an arbitrary number of solitons. However, the total
topological charge of the branes on the circle must vanish,
meaning that there should be an equal number of branes and anti-branes.

We now seek the equilibrium state for such a system.
In the absence of gravity, the dynamics is governed by the following
three types of interaction:
1) a short range attraction between the kink and the anti-kink,
with the strength set by $\lambda$;
2) an analogous repulsion between solitons
set by $a$; and finally, 3) an attraction between a soliton and
the kink (or the anti-kink) governed by $b$.
The latter interaction is attractive because $b>0$, and therefore it
tends to minimize the energy by bringing zeros of $\phi$ and $\cos(\theta)$
close by. Due to this attraction, the solitons form bound states with
either the kink or the anti-kink. As a result, there arise competing forces
between the soliton-kink and soliton-anti-kink bound states thanks
to the soliton-soliton repulsion and the kink-anti-kink attraction. The final
balance depends on the choice of parameters. If the repulsion takes over,
(e.g. for $\lambda^2 v^4 \ll a$) the kink and anti-kink will be stabilized
at the opposite poles of the circle.
This simple example demonstrates how a brane-anti-brane system may become
stable in the presence of extra bulk states.

A more realistic example where the stability of the Manyfold has topological
origin can be constructed as follows.
Let a 3-brane reside in a space with $N$ transverse
dimensions, and let $\phi$ be a real scalar localized on the brane.
If its potential has two disconnected degenerate minima,
such as in Eq. (\ref{pot}), selecting a vacuum will spontaneously break
the symmetry $\phi \rightarrow -\phi$. Therefore there will
exist domain walls residing on the 3-brane, which separate
the vacua where $\phi =\pm v$. Now suppose that the
3-brane is folded precisely at the location of one such domain wall.
The two sides of the fold have the scalar field in a different vacuum.
At large distances, the two folds will attract each other due to
a massless exchange through the bulk, giving
\begin{equation}
 \sim {1 \over r^{N-2}},
\end{equation}
which is dangerous since it could collapse the folds onto each other.
However, the collapse is incompatible with $\phi$
having the opposite expectation values
on the two folds. This gives rise to a repulsive short-range potential
between the folds. Its origin can be deduced as follows: to
bring the two folds together, $\phi$ has to go through zero at each
point of one of the folds. But this costs energy
$\sim \lambda^2 v^4$, which can be made arbitrarily high by
choosing the parameters of the theory.
Thus, we expect that at least in a portion of the parameter space, the
two folds will be stabilized at some finite
distance\footnote{Of course in reality the
``stuff'' that the 3-brane is made of may become important. For example,
one can imagine the situation where the brane annihilation can begin before
the repulsion sets in. However, concrete examples suggest that the
parameters can always be found such that the repulsion can balance
the attraction through the bulk.}.

Additional examples can be constructed along similar lines.
In string theory, for example, there
are non-BPS configurations \cite{sen},
and they are stable in certain cases.
These arguments show that a Manyfold universe may be stable despite the
tendencies towards self-annihilation discussed
above.

A folded brane is automatically
a non-BPS state, breaking all the supersymmetries of the theory.
This remains true for D-branes as well as for the field theoretic solitons.
The signature of the observed SUSY breaking in the brane spectrum,
can be illustrated again by the toy kink-anti-kink brane model above.
For simplicity consider a kink domain wall in a $3 + 1$
dimensional Universe. We define the model such that a
straight wall is a BPS state. This can be achieved for example
in models with a broken discrete $R$-symmetry, e.g. with the
superpotential \cite{ds}
\begin{equation}
W = {\lambda \over 3}\phi^3 + \mu^2\phi.
\end{equation}
This model contains a BPS domain wall, where the supersymmetries
emerge because the background admits a central extension
of the $N=1$ SUSY algebra with a central charge which is a difference of
the $vev$s of the superpotential at infinities on different sides of the wall
\cite{ds}:
\begin{equation}
Q_{central} = 2 [W(+\infty) - W(-\infty)].
\label{cQ}
\end{equation}
This guarantees that $1/2$ of the original supersymmetry is unbroken,
since the central charge exactly cancels the brane tension
in the superalgebra
\begin{equation}
\{Q_{\alpha},Q_{\beta}\} = P_{\alpha\beta} + Q_{central} \Sigma_{\alpha\beta},
\end{equation}
where $\Sigma_{\alpha\beta}$ is proportional to the area tensor.
Thus, there is an exact Fermi-Bose degeneracy in the world-volume theory
of the brane.

Now, it is easy to see that a brane-anti-brane system in this theory
cannot be a BPS state: the central charge vanishes due to
trivial boundary conditions which reflect the topology of the
brane-anti-brane configuration\footnote{If we consider configurations
with say an excess anti-brane number, the boundary conditions
will be again non-trivial. However, supersymmetry can not be restored, since the
balance between the central charge and the multi-brane tension is violated.
In other words, one can not restore supersymmetry by adding more branes.}.
As a result, the
Fermi-Bose degeneracy in the brane world-volume theory is lifted.
For instance, if branes are stabilized at a finite distance,
there is only one massless scalar, corresponding to the center-of-mass
motion, but two massless chiral fermions, one on each fold.
In this way, an observer living in the Manyfold can never see an unbroken
supersymmetry - she will always see an excess of fermionic degrees of freedom.
Hence the Manyfold universe could be an
interesting source of the observed supersymmetry breaking.
It would be interesting to consider in more detail the phenomenological
consequences of this ``brane-mediated" SUSY breaking scenario.

\section{Conclusions}

The Manyfold is a natural setting providing us with:

\noindent (1) A new dark matter candidate and a novel framework for
structure formation;

\noindent (2) A mechanism for neutrino masses;

\noindent (3) A mechanism for supersymmetry breaking.

\noindent In the limit of infinitely many folds the dark matter
becomes dissipationless and replicates the CDM predictions for
structure formation. For $n_f \sim 100$ folds dissipation becomes
important at the present time and a full study will require a
significant effort. Nevertheless there is a strong possibility
that for some range of $n_f$ (or $n_{BR}/n_{BL}$) and $T_R/T_L$ there will be
dark MACHO formation - in the sub-solar mass range - as well as collapses
leading to an increase in gravity waves.

The mechanisms and phenomena introduced here, apart from neutrino
masses, {\it do not} depend on having very large new dimensions $\sim {\rm mm}$
and TeV scale gravity. As long as the bulk is much larger
than $n_f \times l_{P}$ we can use classical higher-dimensional
gravity as a good effective theory in which the Manyfold lives.
So, even if the string scale is not too far from the old-fashioned
$M_{P}$, these ideas can be applied for dark matter, supersymmetry
breaking and astrophysical implications. The one exception is neutrino
masses, where the correct magnitude is reproduced for $\sim$ TeV gravity
and sub-millimeter dimensions.

\vspace{.1cm}
\noindent {\it Note added: After the completion of this manuscript, two papers
appeared \cite{new} which have some overlap with portions
of our Sec 7.}

\vspace{.1cm}
{\bf Acknowledgements}

We would like to thank Ignatios Antoniadis, Glennys Farrar,
John March-Russell, Massimo Porrati, John Preskill,
Mikhail Shifman, Glenn Starkman,
Matt Strassler, Nathan Seiberg, S-H. Henry Tye and
especially Robert Wagoner for useful discussions.
The work of N.A-H. has been supported in part by the DOE under Contract
DE-AC03-76SF00098, and in part by NSF grant PHY-95-14797.
The work of S.D. and N.K has
been supported in part by NSF Grant PHY-9870115.
The work of G.D. has
been supported in part by David and Lucile Packard
Fellowship.

\end{document}